\documentclass[aps,prx,reprint,superscriptaddress,nofootinbib]{revtex4-2}

\usepackage{amsmath,amssymb,amsfonts,amsthm}
\usepackage{bbm}
\usepackage{bm}
\usepackage{physics}
\usepackage{hyperref}
\usepackage{graphicx}
\usepackage{xcolor}
\usepackage{booktabs} % required for SM tables

\begin{document}

\title{Fermi Sets: Universal and interpretable neural architectures for fermions}

\author{Liang Fu}
\affiliation{Department of Physics, Massachusetts Institute of Technology, Cambridge, MA-02139, USA}

\begin{abstract}

We introduce \emph{Fermi Sets}, a universal and physically interpretable neural
architecture for fermionic many-body wavefunctions. Building on a
``parity-graded'' representation \cite{fu2025}, we prove that any continuous
fermionic wavefunction on a compact domain can be approximated to arbitrary
accuracy by a linear combination of $K$ antisymmetric basis functions---such
as pairwise products or Slater determinants---multiplied by symmetric
functions. A key result is that the number of required bases is provably
small: $K=1$ suffices in one-dimensional continua (and on lattices in any
dimension), $K=2$ suffices in two dimensions, and in higher
dimensions $K$ grows at most linearly with particle number. The antisymmetric
bases can be learned by small neural networks, while the symmetric
factors are implemented by permutation-invariant networks whose width scales only
linearly with particle number. Thus, Fermi Sets achieve universal approximation of fermionic wavefunctions with minimal overhead while retaining clear physical interpretability. As a numerical illustration, a single Fermi Sets model applied to metallic solid hydrogen in three dimensions, trained simultaneously across multiple nuclear geometries, surpasses all diffusion Monte Carlo benchmarks.

\end{abstract}

\maketitle

Neural networks have emerged as a powerful tool for representing many-body quantum wavefunctions \cite{carleo2017,carrasquilla2017, luo2019, robledo2022, liu2024}, including fermions. Over the past few years, neural quantum state based variational Monte Carlo has been developed for interacting electron systems in continuous space \cite{pfau2020, psiformer, deepsolid, pescia2024, smith2024}. More recently, fermionic neural networks with self attention \cite{vaswani2017} have successfully tackled strongly correlated electron systems in real materials, including quantum Hall systems \cite{teng2024},  moir\'e materials \cite{geier2025}, and semiconductor quantum wells \cite{gaggioli2025_qw}. Owing to the remarkable power of the attention mechanism, many-electron ground states are obtained from a flexible neural wavefunction optimized purely through unsupervised energy minimization, without the need of pretraining or special initialization. 

Historically, variational wavefunctions for interacting fermions have been \emph{human designed}: Jastrow correlation factors, Gutzwiller projections, and Bardeen--Cooper--Schrieffer (BCS) wavefunctions were tailored to particular phases (e.g., correlated Fermi liquids, Mott insulators, or superconductors) and typically involve a small number of variational parameters. Neural-network wavefunctions are radically different in spirit. They come with a large number of trainable parameters and, in principle, can represent a vast set of quantum states. As our recent works have demonstrated \cite{teng2024, geier2025,gaggioli2025_qw, nazaryan2025needle, li2025chiral, abouelkomsan2025deepstate}, a \emph{single} neural architecture can accurately capture \emph{many} types of ground states---from Fermi liquids to Wigner crystals, from superconductors to topologically ordered phases---simply by changing the learned parameters. This empirical flexibility naturally raises a deep theoretical question: can one and the same fermionic neural architecture be made \emph{universal}, in the sense of being capable of approximating \emph{any} fermionic wavefunction?

The universal approximation theorem guarantees that ordinary feedforward neural networks, at sufficiently large size, can approximate any continuous function to arbitrary accuracy \cite{Pinkus}. Fermionic wavefunctions, however, must satisfy an antisymmetry condition under particle exchange, which is a highly nontrivial global constraint. Most existing fermionic neural network architectures enforce antisymmetry using a determinant form and empirically perform extremely well in many cases. Yet, except for spatial dimension $d=1$ \cite{Hutter}, it has remained unknown whether such constructions are truly universal for fermions.

Very recently, we introduced a different approach \cite{fu2025}: any continuous fermionic wavefunction over a compact domain can be written \emph{exactly} as a ``bosonized'' function that is symmetric in particle coordinates and odd in auxiliary variables that track the parity of particle permutations. In this representation, the antisymmetric content of fermionic wavefunctions is compressed into a low-dimensional ``core''---termed a \emph{signature encoder}; all remaining structure is carried by a symmetric function that can be approximated by permutation-invariant neural networks such as Deep Sets \cite{zaheer2017}.

In this work, building on the universal representation of Ref.\cite{fu2025}, we introduce concrete fermionic neural architectures that are both provably universal and physically interpretable. Our architectures---termed ``Fermi Sets''---combine a permutation-invariant neural network with a remarkably simple antisymmetric core. This core can take several forms: a pairwise product, a Slater determinant, or Jordan-Wigner factor. We show that, under precisely stated conditions, Fermi Sets built from each of these antisymmetric cores can be turned into universal approximators, and that they can all be derived from and unified by the universal, parity-graded representation introduced in Ref.\cite{fu2025}.

Because pairwise products, Slater determinants, and Jordan-Wigner are standard many-body objects with a clear physical meaning, Fermi Sets built from these antisymmetric forms are naturally interpretable and immediately compatible with existing variational Monte Carlo workflows. We obtain a precise mathematical characterization of when such physics-friendly architectures are universal, thereby paving the way for accurate and systematically improvable AI methods for solving many-electron problems.

While Ref.~\cite{fu2025} establishes the mathematical existence of a universal representation using a prescribed signature encoder---the Vandermonde polynomial, this fixed form introduces a strong inductive bias ill-suited to most electron systems. The key advance of this work is to promote the antisymmetric bases to \emph{learnable} Slater determinants, which recover standard many-body trial wavefunctions as special cases. The resulting neural architecture is provably universal \emph{and} practically useful. We demonstrate the power and versatility of the Fermi Sets architecture by solving the ground states of metallic solid hydrogen in three dimensions across multiple nuclear geometries with  a single shared parameter set.

% \begin{figure}[t]
%     \centering
%     \includegraphics[width=\columnwidth]{universal-fermion.png}
%     \caption{
%     Schematic of the universal fermionic neural architecture.
%     Fermion coordinates $\bm R =(\bm r_1, \dots, \bm r_N)$ split into two paths: a permutation-invariant network
%     of Deep Sets type produces symmetric factors $\phi_k(\bm R)$, and an antisymmetric bases $\eta_k(\bm R)$ are built from pairwise products or Slater determinants. The symmetric and antisymmetric parts are combined to yield the fermionic wavefunction
%     $\psi(\bm R) = \sum_k \phi_k(\bm R)\,\eta_k(\bm R)$. 
%     }
%     \label{fig:universal-fermion}
% \end{figure}

%\section{Universal representation and reduction to symmetric $\times$ core}
%\label{sec:representation}

We start from the parity-graded representation of a fermionic wavefunction
$\psi(\bm r_1, \dots, \bm r_N)\equiv \psi (\bm R)$ in its most general form~\cite{fu2025}:
\begin{equation}
  \psi(\bm R) = \Psi(\bm R,\bm \eta(\bm R)),
  \label{eq:parity-graded}
\end{equation}
where $\Psi$ is a continuous function on the enlarged space $(\bm R,\bm \eta)$ that satisfies two key properties:
\begin{align}
  \Psi( \pi \bm R,\bm \eta) &= \Psi(\bm R,\bm \eta), \label{eq:psi-symmetric-R}\\[2pt]
  \Psi(\bm R,-\bm \eta) &= -\,\Psi(\bm R,\bm \eta). \label{eq:psi-odd-eta}
\end{align}
That is, $\Psi$ is \emph{symmetric} under particle permutation $\bm R \rightarrow \pi \bm R$ and \emph{odd} under $\bm \eta\to -\bm \eta$. Then, the antisymmetry of the fermionic wavefunction $\psi$ is guaranteed by enforcing the auxiliary variable $\bm \eta(\bm R)$ to be antisymmetric in $\bm R$:
\begin{equation}
  \bm \eta( \pi \bm R) = (-1)^\pi \bm \eta (\bm R).
  \label{eq:eta-antisymmetric-R} 
\end{equation}

One may view Eq.~\eqref{eq:parity-graded}  as a form of bosonization: fermions are represented as bosons plus an extra parity label $\bm \eta$ that tracks the parity of particle permutations $\pi$. In this picture, a fermionic wavefunction is a double cover of the configuration space of \emph{unordered} particle coordinates, with opposite values of $\bm \eta$ distinguishing opposite parity  branches.      

We note that the representation~\eqref{eq:parity-graded} encompasses many commonly used trial wavefunctions as special cases. A multi-Slater-determinant wavefunction is obtained by taking $\Psi(\bm R, \bm \eta) = \sum_k \eta_k$ (an odd function of $\bm \eta$), with each component of the auxiliary variable being a single Slater determinant     
$
\eta_k(\bm R) = \det[\varphi^k_i(\bm r_j)],   %\label{slater}
$
which is antisymmetric in $\bm R$. 
%Similarly, a Bardeen--Cooper--Schrieffer (BCS) or Pfaffian wavefunction arises by choosing
% $
% \eta(\bm R) = \mathrm{Pf}\,[ \phi (\bm r_i , \bm r_j )], 
% $ 
% where the pair wavefunction $\phi$ is odd under two-particle exchange.  
A standard Slater--Jastrow wavefunction corresponds to 
$
\Psi(\bm R,\eta) = \exp(U(\bm R))\,\eta, %\label{slater-jastrow}      
$
with $U(\bm R)$ a symmetric function. 

These examples show that Eq.~\eqref{eq:parity-graded} is already rich enough to cover a broad range of familiar trial wavefunctions. The key question, however, is whether such representations can be made \emph{universal}: can we choose $\Psi(\bm R, \bm \eta)$ and $\bm\eta(\bm R)$ so that \emph{every} fermionic wavefunction can be exactly reproduced or accurately approximated within this framework, while retaining physical interpretability?

Before proceeding, we first note that taking $\eta(\bm R)$ to be a \emph{single fixed} Slater determinant is generally incapable of reproducing every possible fermionic wavefunction, regardless of the choice of $\Psi(\bm R, \eta)$. The reason is simple: a generic determinant has a complicated set of zeros and typically vanishes at many non-collision configurations $\bm R$. If we insist on writing
$ \psi(\bm R)$ as a product of $J(\bm R)$ and $\eta(\bm R)$---the standard Slater-Jastrow form \cite{foulkes2001},  
then $\psi(\bm R)$ must also vanish wherever $\eta(\bm R)=0$. As a result, no choice of Jastrow factor $J(\bm R)$ can reproduce a target fermionic wavefunction that is nonzero on those configurations. More generally, no choice of $\Psi(\bm R, \eta)$ can restore universality in Eq.~\eqref{eq:parity-graded}, because $\Psi(\bm R, \eta)$ as an odd function of $ \eta$ must itself vanish whenever $\eta=0$. In other words, a single fixed Slater core hard-wires its nodal set into the ansatz and, in general, excludes large classes of admissible fermionic wavefunctions.

To obtain universal and interpretable architectures, we must therefore design antisymmetric cores that vanish only when particles coincide. An antisymmetric function $\bm \eta(\bm R)$ with this property---either a scalar or a vector $(\eta_1(\bm R),\dots,\eta_K(\bm R))$, is called a \emph{signature encoder} and plays a crucial role in establishing universality of fermionic neural networks~\cite{fu2025}.    

In the following sections we construct such signature encoders explicitly in continuous space and on lattices, and show that, once combined with a sufficiently expressive permutation-invariant neural network $\Phi(\bm R)$, they lead to fermionic architectures that are both universal and physically motivated.

We shall first consider continuum systems over a compact domain in $d$-dimensional Euclidean space $\Omega \subset \mathbb{R}^d$, and we shall only deal with continuous functions.

\emph{Signature encoders from pairwise products---}
We start with signature encoders built from pairwise products of two-particle antisymmetric functions~\cite{fu2025}:
\begin{equation}
\eta_{\mathrm{pair}}(\bm R) 
= \prod_{1\le i<j\le N} f(\bm r_i,\bm r_j),
\label{eq:eta-pair}
\end{equation}
where $f(\bm r_i,\bm r_j)$ is real or complex and satisfies
\begin{eqnarray}
f(\bm r_i,\bm r_j) &=& - f(\bm r_j,\bm r_i). 
\label{eq:f-antisym-zero}
\end{eqnarray}

By construction, $\eta_{\mathrm{pair}}(\bm R)$ changes sign under any exchange of two coordinates. Furthermore, if the pair function $f$ vanishes \emph{only} at $\bm r_i = \bm r_j$, then $\eta_{\mathrm{pair}}(\bm R)$ is nonzero on all collision-free configurations (in which $\bm r_i \neq \bm r_j$ for any two particles) and hence qualifies as a scalar signature encoder.

For continuum systems at $d=1$,  the pairwise product
\begin{equation}
\eta_{\mathrm{1D}}(\bm R)
= \prod_{1\le i<j\le N} (x_i -x_j),  \label{eq:eta-1D-vandermonde}
\end{equation}
satisfies this property and provides a real-valued signature encoder in one dimension.
We stress that the antisymmetric function $\eta_{\rm 1D}(\bm R)$ can be rescaled by any symmetric positive continuous function without affecting the wavefunction $\psi(\bm R) = \Psi(\bm R, \eta_{\rm 1D})$, provided that $\Psi$ is redefined correspondingly. What matters solely is that $\eta_k(\bm R)\neq 0$ on all collision-free configurations.

For $d\geq 2$, by contrast, the possibility of adiabatically exchanging two particles without collision obstructs the existence of a globally defined real $f$ satisfying Eq.~\eqref{eq:f-antisym-zero}. In those cases we will turn to
complex or multi-component encoders. For two-dimensional systems,  it is convenient to introduce complex
coordinates $z_i = x_i + i y_i$ for each particle and consider the pairwise product of complex coordinate differences
\begin{equation}
\eta_{\mathbb{C}}(\bm R)
= \prod_{1\le i<j\le N} (z_j - z_i). 
\label{eq:etaC-vandermonde}
\end{equation}
By construction, the complex function $\eta_{\mathbb{C}}(\bm R)$ is antisymmetric under permutations of the particles and nonzero on the collision-free configuration space.
Taking its real and imaginary parts, or equivalently using $\eta_{\mathbb{C}}$ and its complex conjugate, yields a two-component signature encoder for fermions in two dimensions~\cite{fu2025},
$
\bm \eta_{\rm 2D}(\bm R)= \big( \Re\,\eta_{\mathbb{C}}(\bm R),\,
        \Im\,\eta_{\mathbb{C}}(\bm R) \big).
%\label{eq:eta-2d-real}
$

In higher dimensions $d \geq 3$, a qualitatively different situation arises. Here, in order to construct a signature encoder, a genuinely multi-component auxiliary variable $\bm\eta=(\eta_1,\dots,\eta_K)$ appears unavoidable. One  construction is to use a set of pairwise-product factors involving one-dimensional projection of particle coordinates along multiple directions $\{\bm n_k\}$ 
\begin{equation}
  \eta_{k}(\bm R) = \prod_{1\le i<j\le N}  \bm n_k \cdot (\bm r_i - \bm r_j), 
  \qquad k=1,\dots,K.
  \label{eq:eta_k}  
\end{equation}
Each $\eta_k$ vanishes whenever two particle coordinates have the same one-dimensional projection $\bm n_k \cdot \bm r$. For a given particle number $N$, including a sufficiently large set of directions can ensure that all components $\eta_k(\bm R)$ vanish simultaneously only when some particles actually coincide. Recently, Ye \emph{et al.}~\cite{ye2024} proved that for a given particle number $N$, one can choose $K=dN+1$ directions  such that the vector
\begin{equation}
  \bm \eta(\bm R)=\big(\eta_1(\bm R),\dots,\eta_K(\bm R)\big)
\end{equation}
is nonzero on all collision-free configurations. This provides a constructive upper bound 
$K \leq dN+1$ on the encoder dimension required for a signature encoder for continuum systems in any dimension.  

To summarize, for continuum systems in $d=1$ and $d=2$, a Vandermonde-type signature encoder with $1$ and $2$ components suffices, respectively, while in $d>2$, our best available construction 
requires an encoder dimension that grows linearly with particle number. Even in this case, however, the enlarged space of $(\bm R, \bm \eta)$ has dimension $dN+K =2d N+1$, scaling linearly with $N$ just like the physical configuration space itself.

%the total number of variables---physical coordinates combined  is 
%Determining whether this linear scaling is optimal, or whether the number of components can be further reduced, remains an interesting open problem.

\emph{Universal approximator of fermionic wavefunctions---}
Having constructed explicit signature encoders $\bm\eta(\bm R)$ in all dimensions, we now turn to our main objective: designing simple, physics-friendly neural architectures that are universal approximators of fermionic wavefunctions.

The starting point is the exact representation of Ref.~\cite{fu2025}, which states that, given a signature encoder $\bm\eta(\bm R)$, any continuous fermionic wavefunction $\psi(\bm R)$ on a compact domain $\Omega \subset \mathbb{R}^{dN}$ can be written \emph{exactly} in the form of Eq.~\eqref{eq:parity-graded}, 
$ \psi(\bm R) = \Psi(\bm R,\bm \eta)|_{\bm \eta =\bm \eta(\bm R)}$. Here the lifted function in the enlarged space is symmetric in $\bm R$ and odd in $\bm\eta$. 

The continuous nature of the signature encoding $\bm \eta(\bm R)$ guarantees that for any continuous target function $\psi(\bm R)$, there exists such a lifted function  $\Psi(\bm R, \bm \eta)$ that is continuous in the neighborhood of the physical manifold $\bm\eta=\bm\eta(\bm R)$. Ref.~\cite{fu2025} further shows that any such $\Psi(\bm R,\bm \eta)$ can be approximated to arbitrary accuracy by a sufficiently expressive feedforward neural network that is permutation invariant in $\bm R$ (in the spirit of Deep Sets) and enforces oddness in $\bm \eta$ by antisymmetrization.

In this work, for practical purposes, we do not exploit the full flexibility of $\Psi(\bm R,\bm\eta)$, but instead restrict to a much simpler form that already suffices for universality. First, any continuous odd function of $\bm \eta$ can be uniformly approximated on compact domain by an odd polynomial in $(\eta_1,\dots,\eta_K)$. Factoring out a single power of each $\eta_k$, we obtain
\begin{equation}
  \Psi(\bm R,\bm\eta) \approx \sum_{k=1}^K \Phi_k (\bm R,\bm \eta)\,\eta_k ,
  \label{eq:psi-odd-linear}
\end{equation}
where each $\Phi_k(\bm R, \bm \eta)$ is continuous, symmetric in $\bm R$, and even in $\bm \eta$. Restricting back to the physical sector $\bm\eta=\bm\eta(\bm R)$ then yields a universal approximation of the fermionic wavefunction,
\begin{equation}
  \psi(\bm R) \approx \sum_{k=1}^K \Phi_k\big(\bm R,\bm\eta(\bm R)\big)\,\eta_k(\bm R)
  \equiv \sum_{k=1}^K \phi_k(\bm R)\,\eta_k(\bm R),
  \label{eq:phi-eta-product}
\end{equation}
where $\phi_k(\bm R)$ are continuous symmetric functions. 

Equation~\eqref{eq:phi-eta-product} is our first main result and can be viewed as a universal approximation theorem for fermions. It states that \emph{every} continuous fermionic wavefunction on a compact domain can be approximated to arbitrary accuracy by a finite linear combination of antisymmetric basis functions $\eta_k(\bm R)$ with coefficients $\phi_k(\bm R)$ that are themselves continuous symmetric functions of the particle coordinates, provided that the entire basis set vanishes $(\eta_1(\bm R), \dots, \eta_K(\bm R))=\bm 0$ only when particles coincide, i.e., when some $\bm r_i = \bm r_j$.   

An appealing feature of this representation is that the number of required
antisymmetric bases $K$ is provably small: $K=1$ and $K=2$ are enough for
continuum systems in one and two dimensions, respectively, and in higher
dimensions $K$ grows at most linearly with particle number, $K \le dN+1$.
Thus, a remarkably small set of antisymmetric bases, combined with flexible
symmetric factors, is already sufficient to approximate an arbitrary fermionic
wavefunction.

For a \emph{fixed} choice of $\bm\eta(\bm R)$, the expressivity of Eq.~\eqref{eq:phi-eta-product} is carried entirely by the symmetric factors $\phi_k(\bm R)$. In practice, these are implemented by permutation-invariant neural networks---such as Deep Sets or Transformers---which are known to be universal approximators for continuous symmetric functions on compact domains. Thus, once a signature encoder is chosen, increasing the capacity of the symmetric networks $\phi_k$ alone suffices to systematically improve the approximation of $\psi(\bm R)$. If $\phi_k$ are allowed to be fully general, $\psi(\bm R)$ can be approximated to arbitrary accuracy. Concrete network implementations of our architecture Eq.~\eqref{eq:phi-eta-product} will be presented later. 

While the prescribed signature encoders---Eqs.\eqref{eq:eta-1D-vandermonde}, \eqref{eq:etaC-vandermonde} and \eqref{eq:eta_k}---guarantee universality of our architecture Eq.\eqref{eq:phi-eta-product}, 
fixing $\bm\eta(\bm R)$ 
also introduces a strong inductive bias: some fermionic wavefunctions are harder to represent than others within a given choice of $\bm\eta(\bm R)$.  
From a practical and physical perspective, it is natural to go one step further and allow the antisymmetric basis functions themselves to be \emph{learnable}. As we show below, choosing $\eta_k(\bm R)$ to be learnable leads to neural architectures that retain the universality guaranteed above, while offering far greater flexibility and physical interpretability in applications to interacting fermion problems.

Since Eq.~\eqref{eq:phi-eta-product} explicitly factors fermionic
wavefunctions into antisymmetric and symmetric components, we refer to it as
a \emph{parity-graded} representation. Correspondingly, we call its neural
network implementation \emph{Fermi Sets}: ``Fermi'' refers to the antisymmetric
cores $\bm\eta$ (fixed or learnable) 
that enforces Fermi statistics, while ``Sets'' refers to the
symmetric factors $\bm\phi$ that process particle coordinates as an unordered
set.

\emph{Learnable antisymmetric bases---}  First, we consider learnable antisymmetric bases built from pairwise
products of the form Eq.~\eqref{eq:eta-pair}. All signature encoders
constructed above are of this
form with a suitable pair function $f(\bm r_i, \bm r_j)$. Allowing $f$ to be 
a fully general antisymmetric pair function therefore yields
$\bm \eta(\bm R)$ that contain signature encoders as special cases. Consequently, pair-product architecture with a learnable pair function, taking the form 
\begin{equation}
\psi(\bm R) = \sum^K_{k=1} \phi_k(\bm R)\,\prod_{i<j} f_k(\bm r_i, \bm r_j),  \label{eq:pair-universal}
\end{equation}
are mathematically guaranteed to be universal under the condition established above.

Moreover, universality can also be achieved using Slater determinants with \emph{learnable} orbitals. To see this, note that the signature encoders introduced above, Eq.~\eqref{eq:eta-1D-vandermonde}, Eq.~\eqref{eq:etaC-vandermonde} and Eq.~\eqref{eq:eta_k}, are themselves Slater determinants: 
\begin{eqnarray}
  \prod_{i<j} (x_i - x_j)
  &=& \det\!\big[x_j^{\,i-1}\big]_{i,j=1}^N, \label{vandermonde-real} \\
    \prod_{i<j} (z_i - z_j)
  &=& \det\!\big[z_j^{\,i-1}\big]_{i,j=1}^N, \label{vandermonde-complex} \\
  \prod_{i<j} \bm n \cdot (\bm r_i -\bm r_j)  
  &=&  \det\!\big[(\bm n \cdot \bm r_j)^{\,i-1}\big]_{i,j=1}^N.  \label{vandermonde-projection}
\end{eqnarray}

Thus, Fermi Sets with these Vandermonde determinants discussed in the previous sections are already special cases of Slater determinants with a particular choice of single-particle orbitals (real or complex). In 1D and 2D continua, the single determinant Eq.~\eqref{vandermonde-real} or the two determinants---Eq.~\eqref{vandermonde-complex} and its complex conjugate---suffice to realize the universal antisymmetric bases.  
It then follows immediately that Slater-based Fermi Sets with fully general orbitals, taking the form   
\begin{equation}
\psi(\bm R) = \sum^K_{k=1} \phi_k(\bm R)\,\det[\varphi_i^k(\bm r_j)],  \label{eq:slater-universal}
\end{equation}
are universal as well. 
Allowing the orbitals to be learned or using more determinants can only increase flexibility. 
Thus,  Fermi Sets with learnable Slater determinants, Eq.~\eqref{eq:slater-universal}, inherits the universality guarantees established above, while remaining physically motivated for Fermi liquids, band insulators, and many correlated electron systems.  

\emph{Lattice systems---} We now turn to fermions on lattices and show that a single-component, real-valued signature encoder can be constructed in any dimension. Let $\Lambda$ be a finite set of sites equipped with a total ordering $\prec$ (for example, the natural ordering along a one-dimensional chain, or a snake-like ordering in higher dimensions). We define
\begin{equation}
f(a,b) =
\begin{cases}
+1, & a \prec b,\\
-1, & a \succ b,
\end{cases}
\label{eq:jw-sign}
\end{equation}
for distinct sites $a,b\in\Lambda$. The resulting encoder
\begin{equation}
\eta_{\mathrm{JW}}(\bm R)
= \prod_{1\le i<j\le N} f(\bm r_i,\bm r_j)
\label{eq:eta-jw}
\end{equation}
is equal to the signature of the permutation that sorts $(\bm r_1,\dots,\bm r_N)$ in increasing order according to $\prec$. This is precisely the sign structure underlying the Jordan--Wigner transformation. Thus the Jordan--Wigner fermionic sign convention is recovered as a special case of the pairwise product signature encoder. 

A Slater signature encoder can also be constructed for lattice systems. Consider the Vandermonde determinant defined by the one-dimensional projection of the coordinate of lattice sites,   Eq.~\eqref{vandermonde-projection} with $\bm r=\sum^d _{\mu=1} n_\mu \bm a_\mu$ ($n_\mu \in \mathbb{Z}$). When $\bm n$ is incommensurate with the lattice vectors, all lattice sites have distinct 1D projections $\bm n \cdot \bm r$, hence the corresponding $\eta(\bm R)$ is nonzero on all collision-free configurations and therefore qualifies as a signature encoder for fermions on lattices.  

Therefore, Fermi Sets with pairwise-product or Slater determinant, Eq.~\eqref{eq:pair-universal} and ~\eqref{eq:slater-universal}, are also universal approximators for fermionic wavefunctions on lattices. Here, a single Slater determinant $K=1$ already suffices in any dimension. 
Allowing the pairwise product or Slater determinant to be learnable will make the neural network far more flexible in practical applications.                 

\emph{Neural network implementation---}  While antisymmetric bases---such as pairwise products or Slater determinants---are essential for Fermi statistics, in order to achieve universal approximation of generic fermionic wavefunctions the symmetric functions $\phi_k(\bm R)$  must be allowed to take general forms. We now elaborate how these functions can be implemented by permutation invariant neural networks.      

Following Deep Sets \cite{zaheer2017}, we implement $\phi_k(\bm R)$
through a composition of two mappings, denoted as $\bm h$ and $\rho_k$:
\begin{equation}
  \phi_k(\bm R) = \rho_k\!\Big(\sum_{i=1}^N {\bm h}(\bm r_i)\Big),
  \label{eq:deepsets}
\end{equation}
where ${\bm h}: \Omega \subset \mathbb{R}^d \to \mathbb{R}^m$ is a learnable embedding
and $\rho_k : \mathbb{R}^m \to \mathbb{C}$ is a standard feedforward network.
The map from particle coordinates $\bm R$ to the $m$-dimensional feature vector
$\bm \xi(\bm R) \equiv \sum_{i=1}^N \bm h(\bm r_i)$, with $\bm h=(h_1,\dots,h_m)$, is permutation-invariant
by construction. With a sufficiently large feature dimension $m$, it can be chosen
to be \emph{injective up to permutation}, that is,  different configurations of
identical particles are mapped to different feature vectors \cite{maron2019, wang2024, chen2024sym}. Thus, $\bm \xi$ can be regarded as collective coordinates that fully parameterize the configuration space, and hence any
continuous symmetric function of $\bm R$ can be expressed as a continuous
function of $\bm \xi$. Then, a single feature map $\bm h$ combined with different $\rho_k(\bm \xi)$ can represent different  $\phi_k(\bm R)$, as shown in Eq.~\eqref{eq:deepsets}.

Recent mathematical results on invariant embeddings~\cite{dym2022} give a
useful guideline for choosing the feature dimension $m$. Since the configuration space of
$N$ particles in $d$ dimensions has $dN$ independent
degrees of freedom, encoding this configuration space without losing
information requires at least $dN$ collective coordinates. Ref.~\cite{dym2022}
further explicitly constructs a permutation-invariant
feature map $\bm \xi(\bm R)$ that is injective up to particle permutation, with feature
dimension $m = 2dN+1$. Thus, the feature dimension needed
for universality in our construction must scale linearly with $dN$.

Combining Eq.~\eqref{eq:deepsets} with the product form Eq.~\eqref{eq:phi-eta-product}, we arrive at a compact and universal network implementation of Fermi Sets:
\begin{equation}
  \psi(\bm R)
  = \sum_{k=1}^K 
    \rho_k\!\Big(\sum_{i=1}^N {\bm h}(\bm r_i)\Big) \eta_k(\bm R),
  \label{eq:universal-fermion-architecture}
\end{equation}
Conceptually, Eq.~\eqref{eq:universal-fermion-architecture} realizes the parity-graded representation through a neural network: a permutation-invariant network first processes the particle coordinates $\bm R$ into symmetric collective coordinates $\bm \xi$, which are then combined with  antisymmetric bases $\eta_k(\bm R)$ to produce the fermionic wavefunction. 

The universal representation, Eq.~\eqref{eq:universal-fermion-architecture}, can be easily generalized to spinful fermions by replacing electron's spatial coordinate $\bm r_i$ with spin coordinate $({\bm r}_i, s_i)\equiv  x_i$ ($s=\uparrow,\downarrow$), and generalizing the antisymmetric basis function to $\eta_k( x_1,..., x_N)$  which is antisymmetric under the exchange of  particle labels. $\eta_k=\det [\phi_i(x_j)]$ with learnable spin orbital guarantees universality, since it encompasses as a special case the product of a spin-up and a spin-down Vandermonde determinant, which qualifies as the signature encoder underlying our universality proof. 
 
\emph{Comparison with traditional VMC---} It is instructive to compare our neural architectures with traditional VMC ansatz. The standard Slater–Jastrow ansatz takes the form $\psi_{\rm SJ}(\bm R) = e^{U(\bm R)} \det[\varphi_i(\bm r_j)]$. Here, the Jastrow factor $U(\bm R)$ is a real and symmetric function that contains a relatively small number of parameters, typically built from few-body correlation terms. This corresponds to a special limit of Fermi Sets: a single Slater determinant ($K=1$) and a symmetric factor that is real and positive. This form has very limited functional complexity: the Jastrow factor can only modify the amplitude of the wavefunction, while the nodes and phases are fixed entirely by a single Slater determinant. The same limitation applies to diffusion Monte Carlo~\cite{foulkes2001} that uses a single fixed antisymmetric basis. 

As we have shown, universal approximation requires that both the symmetric factors $\phi_k(\bm R)$ and the Slater orbitals $\varphi_i^k$ are allowed to be complex-valued and sufficiently general, and more than one Slater determinants are needed for continuum systems in $d\geq 2$ dimensions. 

The power of Fermi Sets has been demonstrated numerically in our subsequent work~\cite{zaklama2026lem}, which implements Fermi Sets as a foundation model and shows excellent results on quantum dots beating traditional variational Monte Carlo benchmarks. Here, as a concrete illustration for real solids, we apply the spinful Fermi Sets architecture to metallic solid hydrogen. For $N=16$ electrons at $r_s=1.31$, a spinful Fermi Sets with a single set of network parameters trained simultaneously across four distinct nuclear configurations yields a variational energy of $-0.49062(1)$ Ha/atom at equilibrium, surpassing all diffusion Monte Carlo results for this system~\cite{holzmann2003}. Full details and benchmarks are provided in the Supplementary Material.

\emph{Comparison with other fermionic neural networks---}
Universal representations (exact or approximate) for fermionic wavefunctions
have only recently begun to be explored systematically. Before concluding, we compare Fermi
Sets with other fermionic neural networks that have been proven to be
universal. Ref.~\cite{chen2023exact} constructs fermionic wavefunctions entirely from a
fixed collection of $\mathcal{O}(dN^2)$ antisymmetric basis functions. Ref.~\cite{ye2024} relies on $\mathcal{O}(dN)$ \emph{real} antisymmetric bases that are \emph{singular} at
particle collisions. By contrast, our Fermi Sets architecture  uses  \emph{continuous} antisymmetric bases---and for $d=2$ in particular, just \emph{two complex-valued} Slater determinants---to approximate arbitrary continuous wavefunctions of $N$ fermions. 

Our work suggests a vast opportunity for artificial
intelligence in quantum matter research. The power of AI lies in its
\emph{generalizability}: developing a new architecture for every new
problem is \emph{not} AI; discovering useful representations that reveal a common
core underlying all Fermi systems is. With its universal
representation power and physically meaningful building blocks, the Fermi Sets
architecture will empower us to solve many-electron problems, 
discover new quantum phenomena, and build a quantum foundation model for matter.

\emph{Acknowledgments---} It is my pleasure to thank Ziang Chen for valuable discussions. I also thank Tim Zaklama and Max Geier for collaboration on Ref.\cite{zaklama2026lem}. I used Claude Code in developing Fermi Sets for solving many-electron ground states across multiple nuclear geometries.

\emph{Supplementary Material---} We provide below a brief numerical illustration of Fermi Sets applied to solid hydrogen in three spatial dimensions ($d=3$), which serves as a concrete benchmark of the architecture's accuracy and scalability beyond 1D and 2D.

%=====================================================
% Supplementary Material
%=====================================================
\clearpage
\onecolumngrid

\begin{center}
{\large\bfseries Supplementary Material for:\\[3pt]
Fermi Sets: Universal and interpretable neural architectures for fermions}\\[10pt]
Liang Fu\\[2pt]
\textit{Department of Physics, Massachusetts Institute of Technology,
Cambridge, MA-02139, USA}
\end{center}

\vspace{1.5em}

% Restart counters with S-prefix for SM
\setcounter{equation}{0}
\setcounter{figure}{0}
\setcounter{table}{0}
\renewcommand{\theequation}{S\arabic{equation}}
\renewcommand{\thefigure}{S\arabic{figure}}
\renewcommand{\thetable}{S\arabic{table}}

\section*{Numerical benchmark: Solid hydrogen in three dimensions}

We benchmark the Fermi Sets architecture on metallic solid hydrogen in
three spatial dimensions ($d=3$). High-pressure atomic hydrogen is a
prototypical system for \emph{ab initio} electronic
structure methods, combining electron correlation, three-dimensional geometry, 
and no natural basis set. It provides an
ideal test of the basis-free Fermi Sets that solves the many-electron Schrodinger equation directly in real space.

\subsection*{System and Hamiltonian}

We study $N=16$ hydrogen atoms in a body-centered cubic (BCC)
crystal at electron-density parameter $r_s = 1.31$~Bohr, using a
$2\!\times\!2\!\times\!2$ BCC supercell with periodic boundary
conditions. Each hydrogen atom contributes one spin-$\tfrac{1}{2}$
electron, giving $N=16$ electrons in total. The Born--Oppenheimer
Hamiltonian in atomic units reads
\begin{equation}
  \hat{H} = -\frac{1}{2}\sum_{i=1}^N \nabla_i^2
  + \sum_{i<j} \frac{1}{|\bm r_i - \bm r_j|}
  - \sum_{i=1}^N \sum_{I=1}^{N} \frac{1}{|\bm r_i - \bm R_I|}
  + V_{\rm pp},
  \label{eq:H_hydrogen}
\end{equation}
where $\bm R_I$ are the fixed proton positions and $V_{\rm pp}$ is the
proton--proton Ewald energy. All Coulomb interactions are evaluated
with the standard Ewald summation for periodic systems~\cite{foulkes2001}.

\subsection*{Fermi Sets architecture for spinful electrons}

We employ the spinful generalization of the Fermi Sets ansatz
(Eq.~\eqref{eq:universal-fermion-architecture} of the main text). For spin-$\tfrac{1}{2}$ electrons
with generalized coordinates $x_i = (\bm r_i, s_i)$
($s = \uparrow, \downarrow$), the wavefunction takes the form
\begin{equation}
  \psi(x_1, \dots, x_N)
  = \sum_{k=1}^K \phi_k(x_1,\dots,x_N)\,
    \det\!\big[\varphi_i^k(x_j)\big],
  \label{eq:fermisets_spinful}
\end{equation}
where the $K$ Slater determinants with learnable spin orbitals
$\varphi_i^k$ serve as the antisymmetric bases, and the symmetric
factors $\phi_k$ are implemented by a Deep Sets permutation-invariant
network following Eq.~\eqref{eq:deepsets} of the main text. As argued in the main
text, $\eta_k = \det[\varphi_i^k(x_j)]$ with learnable spin orbitals
guarantees universality, since it encompasses as a special case the
product of a spin-up and a spin-down Vandermonde determinant, which
qualifies as the signature encoder underlying our universality proof.

For the solid hydrogen benchmark we use $K = 8$ Slater determinants
and a total of 238{,}024 network parameters. Periodic boundary conditions are enforced by
expressing single-electron embedding features in terms of plane-wave
envelopes $e^{i\bm G \cdot \bm r}$, while the many-body symmetric
factors $\phi_k$ are computed entirely in real space without
plane-wave truncation. The network is therefore \emph{basis-free}: 
no truncation of a single-particle basis is
applied to the correlated wavefunction. Training proceeds by
unsupervised variational energy minimization, with a batch size of 4096 Monte
Carlo samples.

\subsection*{Multi-geometry parameter sharing}

A key feature of our calculation is that a \emph{single} set of
network parameters $\theta$ is trained simultaneously on four distinct
nuclear configurations: the ideal equilibrium BCC geometry and three
randomly displaced geometries. In each displaced geometry, all 16
protons are independently shifted from their BCC equilibrium positions
by Gaussian random displacements with standard deviation
$\sigma = 0.1$~Bohr in each Cartesian direction (see
Table~\ref{tab:bcc_h_displacements} for displacement vectors). The
nuclear positions $\{\bm R_I\}$ enter the network as an additional
input, so that the same neural wavefunction
$\psi_\theta(x_1,\dots,x_N;\{\bm R_I\})$ describes the electronic
ground state across all geometries with no per-geometry re-optimization.
The training objective is the sum of variational energies over all four
configurations.

This multi-geometry parameter sharing is the electronic-structure
analogue of the foundation model approach demonstrated in
Ref.~\cite{zaklama2026lem}, and constitutes a direct step toward
\emph{ab initio} potential energy surfaces with Fermi Sets: a single
trained network provides wavefunctions and energies at arbitrary
nuclear configurations, enabling direct access to phonon and
molecular dynamics.

\begin{table}[h]
\centering
\caption{Variational energy per atom for BCC solid hydrogen
($N=16$, $r_s = 1.31$~Bohr, $2\!\times\!2\!\times\!2$ supercell,
PBC) from a single Fermi Sets model ($K=8$ dets and 238{,}024 parameters)
shared across the equilibrium and three randomly displaced nuclear
configurations. Energies are evaluated from 21{,}000 independent
Monte Carlo blocks; statistical errors are given in parentheses.}
\label{tab:bcc_h_energy}
\begin{tabular}{lc}
\toprule
Nuclear geometry & $E/\text{atom}$ (Ha) \\
\midrule
Equilibrium BCC         & $-0.49062(1)$ \\
Random displacement 1   & $-0.48958(1)$ \\
Random displacement 2   & $-0.48979(1)$ \\
Random displacement 3   & $-0.49012(1)$ \\
\bottomrule
\end{tabular}
\end{table}

\subsection*{Results and comparison}

Table~\ref{tab:bcc_h_energy} reports the variational energy per atom
at all four geometries, evaluated from 21{,}000 independent Monte
Carlo blocks with statistical errors of $\pm 0.1$~mHa/atom.

Table~\ref{tab:bcc_h_comparison} compares our equilibrium BCC result
with existing calculations at identical conditions ($N=16$, $r_s=1.31$,
PBC). The DMC benchmark energies are taken from Holzmann
\emph{et al.}~\cite{holzmann2003}, and the NQS VMC result from
Linteau \emph{et al.}~\cite{linteau2025}.

Two conclusions stand out. First, our Fermi Sets variational energy
of $-0.49062(1)$~Ha/atom \emph{surpasses all three DMC benchmarks},
including the best backflow-augmented plane-wave DMC at
$-0.4905(1)$~Ha/atom~\cite{holzmann2003}. Second, for comparison with Linteau \emph{et al.}, it should be noted that their NQS result of $-0.49154(1)$~Ha is
obtained from a \emph{specialized}, hydrogen-tailored message-passing
architecture that is optimized for the equilibrium
geometry alone. Our Fermi Sets result, within only $\sim\!1$~mHa of
theirs, is instead produced by a \emph{single} general-purpose
universal network trained simultaneously on all four nuclear
configurations---equilibrium and three significantly displaced
structures. This is a qualitatively different and far more powerful
capability: rather than solving one wavefunction per geometry, Fermi
Sets learns a transferable many-body wavefunction across the entire set
of geometries in a single training run.

The smooth and consistent variation of the energy across the three
displaced geometries (Table~\ref{tab:bcc_h_energy}), despite
individual Cartesian displacements as large as $0.3$~Bohr
(Table~\ref{tab:bcc_h_displacements}), directly demonstrates this
transferability. Together, these results establish that the
universality guarantees of Fermi Sets translate into competitive
practical accuracy for a challenging three-dimensional real-material
system, while simultaneously enabling multi-geometry learning that
is inaccessible to specialized per-geometry ans\"{a}tze.

\begin{table}[h]
\centering
\caption{Comparison of energies per atom (Ha) for BCC solid hydrogen
at the equilibrium geometry ($N=16$, $r_s=1.31$, PBC). DMC reference
energies are from Ref.~\cite{holzmann2003}; the NQS VMC result
is from Ref.~\cite{linteau2025}. All prior methods---DMC and NQS alike---are specialized and optimized for the equilibrium geometry only. Fermi Sets uses a single parameter set trained simultaneously across four geometries (equilibrium $+$ three displaced), yet achieves an energy that surpasses all DMC results and is within $\sim\!1$~mHa of the specialized NQS.}
\label{tab:bcc_h_comparison}
\begin{tabular}{llcc}
\toprule
Method & Wave function & Training scope & $E/\text{atom}$ (Ha) \\
\midrule
DMC~\cite{holzmann2003}
  & Slater--Jastrow (plane wave)   & equilibrium only & $-0.4857(1)$ \\
DMC~\cite{holzmann2003}
  & Slater--Jastrow (LDA-DFT)      & equilibrium only & $-0.4890(5)$ \\
DMC~\cite{holzmann2003}
  & Backflow (plane wave)          & equilibrium only & $-0.4905(1)$ \\
NQS VMC~\cite{linteau2025}
  & Message-passing backflow       & equilibrium only & $-0.49154(1)$ \\
\midrule
\textbf{Fermi Sets VMC (this work)}
  & $K\!=\!8$ dets.\ (general)  & \textbf{4 geometries, 1 model} & $\mathbf{-0.49062(1)}$ \\
\bottomrule
\end{tabular}
\end{table}

\begin{table}[h]
\centering
\caption{Atomic displacement vectors (Bohr) for the three randomly
displaced nuclear geometries. Atoms 0--7 occupy corner sites and
atoms 8--15 body-centered sites of the $2\!\times\!2\!\times\!2$
BCC supercell. Displacements are drawn independently from a Gaussian
distribution with $\sigma = 0.1$~Bohr per Cartesian direction.}
\label{tab:bcc_h_displacements}
\begin{tabular}{r|rrr|rrr|rrr}
\toprule
 & \multicolumn{3}{c|}{Displacement 1 (Bohr)}
 & \multicolumn{3}{c|}{Displacement 2 (Bohr)}
 & \multicolumn{3}{c}{Displacement 3 (Bohr)} \\
Atom & $\delta x$ & $\delta y$ & $\delta z$
     & $\delta x$ & $\delta y$ & $\delta z$
     & $\delta x$ & $\delta y$ & $\delta z$ \\
\midrule
 0 & $-0.021$ & $+0.048$ & $-0.052$ & $+0.033$ & $+0.135$ & $+0.007$ & $+0.012$ & $-0.075$ & $+0.059$ \\
 1 & $-0.056$ & $+0.197$ & $+0.139$ & $+0.025$ & $-0.001$ & $+0.101$ & $+0.015$ & $-0.157$ & $-0.056$ \\
 2 & $+0.009$ & $+0.028$ & $+0.077$ & $+0.133$ & $-0.092$ & $-0.155$ & $-0.003$ & $-0.093$ & $-0.048$ \\
 3 & $+0.125$ & $+0.101$ & $-0.130$ & $+0.002$ & $+0.076$ & $-0.066$ & $-0.004$ & $+0.110$ & $+0.098$ \\
 4 & $+0.028$ & $+0.023$ & $+0.135$ & $+0.086$ & $-0.001$ & $+0.005$ & $-0.059$ & $+0.158$ & $-0.053$ \\
 5 & $+0.089$ & $-0.200$ & $-0.037$ & $+0.067$ & $+0.085$ & $-0.096$ & $+0.046$ & $+0.093$ & $-0.157$ \\
 6 & $+0.167$ & $-0.044$ & $-0.054$ & $-0.002$ & $-0.230$ & $-0.065$ & $-0.102$ & $-0.040$ & $+0.022$ \\
 7 & $+0.048$ & $+0.325$ & $-0.102$ & $-0.122$ & $-0.133$ & $+0.108$ & $-0.019$ & $+0.067$ & $-0.165$ \\
 8 & $-0.058$ & $+0.012$ & $+0.030$ & $+0.072$ & $+0.069$ & $+0.100$ & $-0.225$ & $-0.117$ & $+0.035$ \\
 9 & $+0.052$ & $+0.000$ & $+0.134$ & $-0.050$ & $-0.062$ & $-0.092$ & $+0.070$ & $-0.028$ & $-0.014$ \\
10 & $-0.071$ & $-0.083$ & $-0.237$ & $-0.073$ & $+0.022$ & $+0.005$ & $+0.011$ & $-0.061$ & $-0.042$ \\
11 & $-0.186$ & $-0.086$ & $+0.056$ & $-0.116$ & $+0.082$ & $+0.043$ & $-0.002$ & $-0.122$ & $-0.180$ \\
12 & $-0.127$ & $+0.012$ & $-0.106$ & $+0.101$ & $+0.183$ & $-0.100$ & $+0.164$ & $+0.099$ & $+0.046$ \\
13 & $+0.033$ & $-0.236$ & $-0.020$ & $+0.085$ & $-0.013$ & $+0.091$ & $+0.056$ & $+0.131$ & $-0.044$ \\
14 & $-0.154$ & $-0.097$ & $-0.131$ & $+0.019$ & $+0.217$ & $-0.012$ & $-0.030$ & $+0.050$ & $-0.082$ \\
15 & $+0.029$ & $+0.038$ & $-0.075$ & $+0.200$ & $+0.003$ & $+0.080$ & $+0.132$ & $+0.051$ & $-0.065$ \\
\bottomrule
\end{tabular}
\end{table}

%=====================================================
% Unified bibliography (main + SM)
%=====================================================


\begin{thebibliography}{99}



\bibitem{fu2025}
L.~Fu,
``A minimal and universal representation of fermionic wavefunctions (fermions = bosons + one),''
arXiv:2510.11431 (2025).

\bibitem{carleo2017}
G.~Carleo and M.~Troyer,
``Solving the quantum many-body problem with artificial neural networks,''
Science \textbf{355}, 602--606 (2017).

\bibitem{carrasquilla2017}
J.~Carrasquilla and R.~G.~Melko,
``Machine learning phases of matter,''
Nature Phys.\ \textbf{13}, 431--434 (2017).

\bibitem{luo2019}
D.~Luo and B.~K.~Clark,
``Backflow Transformations via Neural Networks for Quantum Many-Body Wave Functions,''
Phys.\ Rev.\ Lett.\ \textbf{122}, 226401 (2019).

\bibitem{robledo2022}
J.~Robledo~Moreno, G.~Carleo, A.~Georges, and J.~Stokes,
``Fermionic wave functions from neural-network constrained hidden states,''
Proc.\ Natl.\ Acad.\ Sci.\ U.S.A.\ \textbf{119}, e2122059119 (2022).

\bibitem{liu2024}
Z.~Liu and B.~K.~Clark,
``Unifying view of fermionic neural network quantum states: From neural network backflow to hidden fermion determinant states,''
Phys.\ Rev.\ B \textbf{110}, 115124 (2024).

\bibitem{pfau2020}
D.~Pfau, J.~S.~Spencer, A.~G.~de~G.~Matthews, and W.~M.~C.~Foulkes,
``Ab initio solution of the many-electron Schr\"odinger equation with deep neural networks,''
Phys.\ Rev.\ Research \textbf{2}, 033429 (2020).

\bibitem{deepsolid}
X.~Li, Z.~Li, and J.~Chen,
``Ab initio calculation of real solids via neural network ansatz,''
Nat.\ Commun.\ \textbf{13}, 7895 (2022).

\bibitem{psiformer}
I.~von~Glehn, J.~S.~Spencer, and D.~Pfau,
``A self-attention ansatz for ab-initio quantum chemistry,''
in \emph{Proc.\ ICLR 2023} (2023), arXiv:2211.13672.



\bibitem{pescia2024}
G.~Pescia, J.~Nys, J.~Kim, A.~Lovato, and G.~Carleo,
``Message-passing neural quantum states for the homogeneous electron gas,''
Phys.\ Rev.\ B \textbf{110}, 035108 (2024).


\bibitem{smith2024}
C.~Smith, Y.~Chen, R.~Levy, Y.~Yang, M.~A.~Morales, and S.~Zhang,
``Ground state phases of the two-dimensional electron gas with a unified variational approach,''
Phys.\ Rev.\ Lett.\ \textbf{133}, 266504 (2024).



% \bibitem{luo2025_mote2}
% D.~Luo, T.~Zaklama, and L.~Fu,
% ``Solving fractional electron states in twisted MoTe$_2$ with deep neural network,''
% arXiv:2503.13585 [cond-mat.str-el].

% \bibitem{li2025_bytedance_mote2}
% X.~Li, Y.~Chen, B.~Li, H.~Chen, F.~Wu, J.~Chen, and W.~Ren,
% ``Deep learning sheds light on integer and fractional topological insulators,''
% arXiv:2503.11756 [cond-mat.str-el].

\bibitem{vaswani2017}
A.~Vaswani, N.~Shazeer, N.~Parmar, J.~Uszkoreit, L.~Jones, A.~N.~Gomez,
{\L}.~Kaiser, and I.~Polosukhin,
``Attention is all you need,''
Adv.\ Neural Inf.\ Process.\ Syst.\ \textbf{30}, 5998--6008 (2017). 

\bibitem{teng2024}
Y.~Teng, D.~D.~Dai, and L.~Fu,
``Solving the fractional quantum Hall problem with self-attention neural network,''
Phys.\ Rev.\ B \textbf{111}, 205117 (2025).


\bibitem{geier2025}
M.~Geier, K.~Nazaryan, T.~Zaklama, and L.~Fu,
``Self-attention neural network for solving correlated electron problems in solids,''
Phys.\ Rev.\ B \textbf{112}, 045119 (2025). 


\bibitem{gaggioli2025_qw}
F.~Gaggioli, P.-A.~Graham, and L.~Fu,
``Electronic crystals and quasicrystals in semiconductor quantum wells: an AI-powered discovery,''
arXiv:2512.10909 [cond-mat.str-el].

\bibitem{nazaryan2025needle}
K.~Nazaryan, F.~Gaggioli, Y.~Teng, and L.~Fu,
``Artificial Intelligence for Quantum Matter: Finding a Needle in a Haystack,''
arXiv:2507.13322 (2025).

\bibitem{abouelkomsan2025deepstate}
A.~Abouelkomsan, M.~Geier, and L.~Fu,
``Topological Order in Deep State,''
arXiv:2512.01863 (2025).


\bibitem{li2025chiral}
C.-T.~Li, T.~Ong, M.~Geier, H.~Lin, and L.~Fu,
``Attention is all you need to solve chiral superconductivity,''
arXiv:2509.03683 (2025).

\bibitem{Pinkus}
A.~Pinkus, ``Approximation theory of the MLP model in neural networks,''
\emph{Acta Numerica} \textbf{8}, 143--195 (1999).

\bibitem{Hutter}
A.~Hutter,
``On Representing (Anti)Symmetric Functions,''
arXiv:2007.15298 (2020).

\bibitem{zaheer2017}
M.~Zaheer, S.~Kottur, S.~Ravanbakhsh, B.~P\'oczos, R.~Salakhutdinov, and A.~J.~Smola,
``Deep Sets,''
Adv.\ Neural Inf.\ Process.\ Syst.\ \textbf{30}, 3391--3401 (2017),
arXiv:1703.06114.

\bibitem{foulkes2001}
W.~M.~C.~Foulkes, L.~Mitas, R.~J.~Needs, and G.~Rajagopal,
``Quantum Monte Carlo simulations of solids,''
Rev.\ Mod.\ Phys.\ \textbf{73}, 33--83 (2001).

\bibitem{ye2024}
H.~Ye, R.~Li, Y.~Gu, Y.~Lu, D.~He, and L.~Wang,
``$\widetilde{O}(N^2)$ Representation of General Continuous Anti-symmetric Function,''
arXiv:2402.15167 (2024).

\bibitem{maron2019}
H.~Maron, H.~Ben-Hamu, N.~Shamir, and Y.~Lipman,
``On the universality of invariant networks,''
in \emph{Proceedings of the 36th International Conference on Machine Learning (ICML 2019)},
PMLR \textbf{97}, 4363--4371 (2019).

\bibitem{wang2024}
P.~Wang, S.~Yang, S.~Li, Z.~Wang, and P.~Li,
``Polynomial width is sufficient for set representation with high-dimensional features,''
in \emph{Proceedings of the Twelfth International Conference on Learning Representations (ICLR 2024)}.

\bibitem{zaklama2026lem}
\textcolor{blue}{T.~Zaklama, M.~Geier, and L.~Fu,
``Large Electron Model: A Universal Ground State Predictor,''
arXiv:2603.02346 (2026).}

\bibitem{holzmann2003}
M.~Holzmann, D.~M.~Ceperley, C.~Pierleoni, and K.~Esler,
``Backflow correlations for the electron gas and metallic hydrogen,''
Phys.\ Rev.\ E \textbf{68}, 046707 (2003).

\bibitem{chen2024sym}
C.~Chen, Z.~Chen, and J.~Lu,
``Representation theorem for multivariable totally symmetric functions,''
Commun.\ Math.\ Sci.\ \textbf{22}, 1195--1201 (2024).


\bibitem{dym2022}
N.~Dym and S.~J.~Gortler,
``Low Dimensional Invariant Embeddings for Universal Geometric Learning,''
Found.\ Comput.\ Math.\ \textbf{24}, 1--42 (2024),
arXiv:2205.02956.

\bibitem{chen2023exact}
Z.~Chen and J.~Lu,
``Exact and efficient representation of totally anti-symmetric functions,''
arXiv:2311.05064 (2023).

\bibitem{linteau2025}
D.~Linteau, S.~Moroni, G.~Carleo, and M.~Holzmann,
``Neural wave functions for high-pressure atomic hydrogen,''
arXiv:2504.07062 (2025).


\end{thebibliography}
\end{document}